\documentclass[aps,prl,twocolumn]{revtex4}
\usepackage{graphics}
\usepackage{graphicx,color}
\usepackage{amsmath,amsthm,amsfonts,amssymb,bm}
\usepackage{epsf}

\begin{document}

\title{Preparing two-atom entangled state in a cavity and probing it via quantum nondemolition measurement}
\author{D. Z. Rossatto}
\author{C. J. Villas-Boas}
\affiliation{Departamento de F\'{\i}sica, Universidade Federal de S\~{a}o Carlos, CEP 13565-905, S\~{a}o Carlos, SP, Brazil}

\begin{abstract}
We propose a probabilistic scheme to prepare a maximally entangled state
between a pair of two-level atoms inside a leaking cavity, without requiring
precise time-controlling of the system evolution and initial atomic state. We
show that the steady state of this dissipative system is a mixture of two
parts: either the atoms being in their ground state or in a maximally
entangled one. Then, by applying a weak probe field on the cavity mode we are
able to distinguish those states without disturbing the atomic system, i.e.,
performing a quantum nondemolition measurement via the cavity transmission. In
this scheme, one has nonzero cavity transmission only when the atomic system
is in an entangled state so that a single click in the detector is enough to
ensure that the atoms are in an maximally entangled state. Our scheme relies
on an interference effect as it happens in electromagnetically induced
transparency phenomenon so that it works out even in the limit of decay rate
of the cavity mode much stronger than the atom-field coupling.

\end{abstract}
\maketitle

The preparation and manipulation of entangled states have attracted much
interest in last years, as they do not have a classical counterpart. These
states are key ingredients for quantum nonlocality tests \cite{Bell} and play
an important role in achieving tasks of quantum computation and communication
\cite{Bennett}, such as quantum cryptography \cite{Ekert}, computers
\cite{Gottesman} and teleportation \cite{Bennett2}. Entangled states can be
prepared either directly by coherent control of unitary dynamics
\cite{Haroche}, as consequence of measurements \cite{Plenio}, or even using a
dissipative process \cite{Beige}. Recently, preparing quantum systems in an
entangled state by dissipative schemes has been actively studied since the
noise, which is always present in the experiments, can be used as a resource
for entanglement generation, avoiding the usual destructive effect on the
quantum system coherence owing to the system-environment interaction.

On the other hand, entanglement quantifiers, such as concurrence
\cite{Wooters} and negativity \cite{Peres}, are not physical observables,
i.e., there are no directly measurable observables, until now, to describe the
entanglement of a given arbitrary quantum state. In general, it is necessary
to perform the quantum state tomography to calculate these entanglement
quantifiers, perturbing the state of the system, although some interesting
methods have been recently proposed to construct direct observables related to
entanglement \cite{Walborn,Mintert,Yu,Yu2,Yu3}. Whereas the authors in Refs.
\cite{Walborn,Mintert,Yu,Yu2} can determine the entanglement when few copies
of the quantum system are available, in Ref. \cite{Yu3} the authors do this by
introducing a probe atom that performs a quantum nondemolition measurement.

Here we propose a probabilistic scheme to prepare a maximally entangled state
between a pair of two-level atoms inside a leaking cavity, without requiring
precise time-controlling of the system evolution or strong atom-field
coupling. The steady state $\left(  \rho_{ss}\right)  $ of this dissipative
process is a mixed state with two parts: one of them describing the
possibility of having both atoms in the ground state $\left\vert
G\right\rangle $ and another one describing the atoms in a maximally entangled
state $\left\vert D\right\rangle $. In both cases the cavity mode is in the
vacuum. From the view point of a single experimental run, $\rho_{ss}$ shows us
that the atomic system can be either in an uncorrelated state or in
a\ maximally entangled one. In this way, if we are able to distinguish both
states without perturbing the atomic system, then we will be able to prepare
it in a maximally entangled state. In fact, we can do this by employing a weak
probe field on the atom-cavity system. As we show bellow, when the atomic
system is in an uncorrelated state the cavity transmission goes to zero,
contrary to the maximum transmission which happens only when the atomic system
is in the maximally entangled state. So, a single click on the detector works
out as a witness of the entanglement generation of the atomic system. On the
other hand, we also show that, if we have a unknown mixed state $\rho$
(between the states $\left\vert G\right\rangle $ and $\left\vert
D\right\rangle $), the average transmission of the atom-cavity system is
exactly equal to the concurrence of the state $\rho$, thus providing a direct
method to measure the degree of entanglement of the atomic system.

\begin{figure}[htbp]
\includegraphics[height=3.2266in,width=2.4284in]{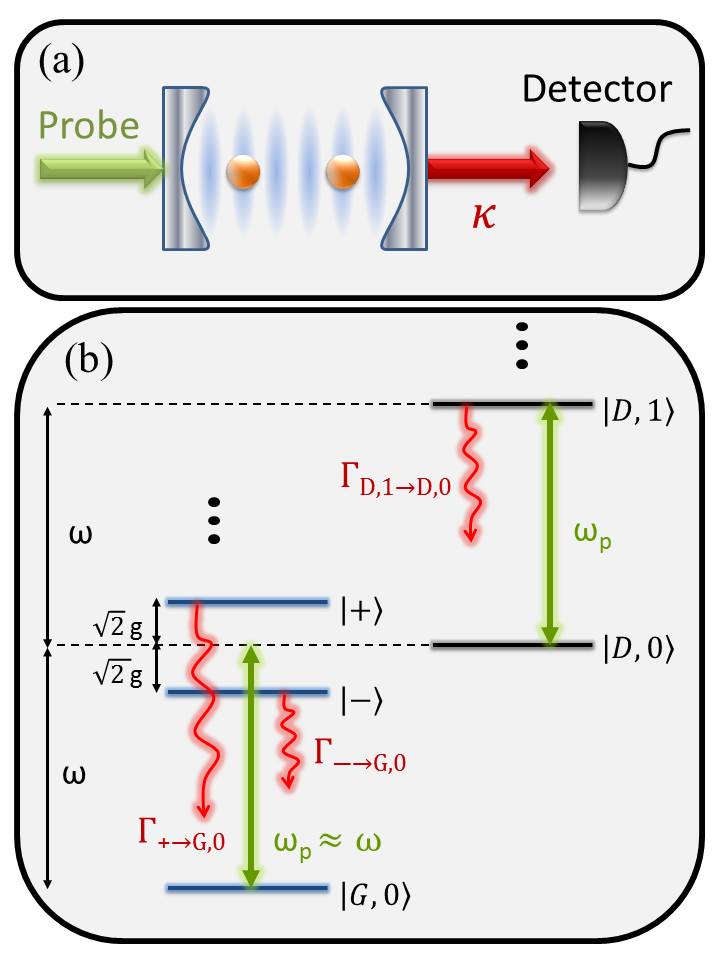}{}
\caption{(color online). (a) Experimental setup. A pair of two-level atoms
inside a leaking cavity. Once the system reaches the steady state the weak
probe field is switched on and the cavity transmission is monitored. (b)
Energy level diagram of the whole system considering the decay rates and the
probe field.}
\label{setup}
\end{figure}

\textit{Model:} Consider a pair of identical two-level atoms ($\left\vert
g\right\rangle _{j}$ = \ ground state, $\left\vert e\right\rangle _{j}$ =
\ excited state) coupled resonantly with a cavity mode with coupling strength
$g$, modeled by Tavis-Cummings Hamiltonian ($\hbar=1$) \cite{Tavis}
\begin{equation}
H=\omega a^{\dag}a+\frac{\omega}{2}S_{z}+g\left(  aS_{+}+a^{\dag}S_{-}\right)
, \label{Hamiltoniano}
\end{equation}
where the cavity mode and the atomic transition are at frequency $\omega$. The
operators $S_{z}\equiv
{\textstyle\sum\nolimits_{j=1}^{2}}
\sigma_{z}^{j}$ and $S_{\pm}=
{\textstyle\sum\nolimits_{j=1}^{2}}
\sigma_{\pm}^{j}$ are the collective spin operators \cite{Dicke} with
$\sigma_{\pm}^{j}=\left(  \sigma_{x}^{j}\pm i\sigma_{y}^{j}\right)  /2$ and
$\sigma_{x,y,z}^{j}$ being the Pauli operators for each atom; $a$ $\left(
a^{\dagger}\right)  $ is the annihilation (creation) operator of the cavity
field. Assuming a leaking cavity at zero temperature, the dynamics of this
system is governed by the master equation \cite{Breuer}

\begin{equation}
\dot{\rho}=-i\left[  H,\rho\right]  +\kappa\left(  2a\rho a^{\dag}-a^{\dag
}a\rho-\rho a^{\dag}a\right)  , \label{masteq}
\end{equation}
with $\kappa$ being the dissipation rate of the cavity mode. The proposed
experimental setup is shown in Fig. \ref{setup}(a).

The spectrum of the system, i.e., the allowed states of it, is given by the
dressed states of $H$
\begin{subequations}
\label{eigenstates}
\begin{align}
&  \left\vert G,0\right\rangle =\left\vert G\right\rangle \otimes\left\vert
0\right\rangle _{c},\label{g}\\
&  \left\vert \pm\right\rangle =\frac{1}{\sqrt{2}}\left(  \left\vert
B\right\rangle \otimes\left\vert 0\right\rangle _{c}\pm\left\vert
G\right\rangle \otimes\left\vert 1\right\rangle _{c}\right)  ,\label{pm}\\
&  \left\vert D,n\right\rangle =\left\vert D\right\rangle \otimes\left\vert
n\right\rangle _{c}, \label{d}
\end{align}
with energies $-\omega$, $\pm\sqrt{2}g$ and $n\omega$, respectively, where
\end{subequations}
\begin{subequations}
\begin{align}
\left\vert G\right\rangle  &  =\left\vert g\right\rangle _{1}\otimes\left\vert
g\right\rangle _{2},\\
\left\vert B\right\rangle  &  =\left(  \left\vert g\right\rangle _{1}
\otimes\left\vert e\right\rangle _{2}+\left\vert e\right\rangle _{1}
\otimes\left\vert g\right\rangle _{2}\right)  /\sqrt{2},\\
\left\vert D\right\rangle  &  =\left(  \left\vert g\right\rangle _{1}
\otimes\left\vert e\right\rangle _{2}-\left\vert e\right\rangle _{1}
\otimes\left\vert g\right\rangle _{2}\right)  /\sqrt{2},
\end{align}
and $\left\vert n\right\rangle _{c}$ is the cavity mode state in the Fock
basis, with $n=0,1$. Here we are considering only the ground, first and second
excited eigenstates of our system once we are interested in the steady state
which is a mixture of those eigenstates. Besides, our scheme requires a weak
probe field which also keeps the cavity field with up to one photon as we will
explain bellow.

The damping of the cavity mode can promote transitions between the eigenstates
of the system whose rates can be obtained through the Fermi gold rule
\cite{Sanders}. As we are considering only the cavity decay, the transition
rate from a higher energy state $\left\vert i\right\rangle $ to a lower one
$\left\vert f\right\rangle $ is given by $\Gamma_{i\rightarrow f}
=\kappa\left\vert \left\langle f\right\vert a\left\vert i\right\rangle
\right\vert ^{2}$. When we take into account the eigenstates of our system it
is easy to see that we have two independent subspaces: $\left\{  \left\vert
G,0\right\rangle \rightarrow\text{ }\left\vert -\right\rangle \text{ or
}\left\vert +\right\rangle \right\}  $ and $\left\{  \left\vert
D,n\right\rangle \right\}  $, i.e., there is no transitions between states
which belongs to distinct subspaces. Therefore, the nonzero transition rates
are $\Gamma_{\pm\rightarrow G,0}=\kappa/2$ and $\Gamma_{D,n+1\rightarrow
D,n}=\kappa$. In the Fig. \ref{setup}(b) is depicted the energy level diagram
of the whole system considering the decay rates and the probe field (frequency
$\omega_{p}$) that will be introduced later.

Owing to the existence of two independent subspaces, for any general initial
state, the steady state of the system is a mixture between the lowest energy
eigenstates of each subspace, i.e.,
\end{subequations}
\begin{equation}
\rho_{ss}=\left(  1-P\right)  \left\vert G,0\right\rangle \left\langle
G,0\right\vert +P\left\vert D,0\right\rangle \left\langle D,0\right\vert ,
\label{rhoss}
\end{equation}
with $P=Tr\left[  \rho\left(  0\right)  \left\vert D\right\rangle \left\langle
D\right\vert \right]  $ being the projection of the initial state on the dark
state $\left\vert D\right\rangle $. This result can be obtained directly from
the Eq. (\ref{masteq}) for $t\rightarrow\infty$ $\left(  \dot{\rho}=0\right)
$. It is important to emphasize that we are not considering atomic damping as
it destroys the entanglement in the steady state so that $\rho(t\rightarrow
\infty)\rightarrow\left\vert G,0\right\rangle \left\langle G,0\right\vert $
for any initial state. As a real two level system always has a spontaneous
decay $\gamma$, our results are valid in a time window defined by $\gamma
t\lesssim1$ and $g^{2}t/\kappa\gg1$ so that we must have $g^{2}/\kappa
\gg\gamma$ \cite{effmasteq}.

From the point of view of a single experimental run, we can see from the
$\rho_{ss}$ that the system can be either in the atomic ground state
$\left\vert G\right\rangle $ or in the entangled state $\left\vert
D\right\rangle $ with probabilities $1-P$ and $P$, respectively. So, there is
a probability of having the atoms in a maximally entangled state. However, a
direct measurement of the atoms would destroy such entangled state. To
circumvent this problem, we must be able to nondestructively measure our
atomic system. We can do that by probing our system with a weak probe field
which allow us to distinguish the atomic states ($\left\vert G\right\rangle $
or $\left\vert D\right\rangle $) through the cavity transmission without
disturbing the atomic system.

To nondestructively measure the system, firstly we must wait until the system
reach its steady state. Then, we apply a weak probe field on the cavity, whose
Hamiltonian is described by
\begin{equation}
H_{p}=\varepsilon\left(  ae^{i\omega_{p}t}+a^{\dag}e^{-i\omega_{p}t}\right)  ,
\end{equation}
with $\varepsilon\ll g$. Here, $\varepsilon$ and $\omega_{p}$ are the strength
and the frequency of the probe field, respectively.

In order to understand how this probe field can provide us information about
the atomic state, firstly we will consider the resonant case, i.e.,
$\omega_{p}=\omega$. If the system is in the $\left\vert D,0\right\rangle $
state, we can see from the Fig. \ref{setup}(b) that the probe field is able to
promote the transition $\left\vert D,0\right\rangle \leftrightarrow\left\vert
D,1\right\rangle $. However, as $\left\vert D\right\rangle $ is a dark state,
it is decoupled from the cavity mode so that the system behaves as an empty
cavity case ($g=0$). In this case, the asymptotic cavity field state is a
coherent field $\left\vert \alpha\right\rangle _{c}=e^{-\left\vert
\alpha\right\vert ^{2}/2}\left(  \left\vert 0\right\rangle _{c}+\alpha
\left\vert 1\right\rangle _{c}+...\right)  $, with $\alpha=-i\varepsilon
/\kappa$. Then, for very weak probe field ($\varepsilon\ll\kappa$) the steady
state of the atom-field system will be given by
\begin{equation}
\left\vert \psi\right\rangle _{ss}^{D}\approx\left\vert D\right\rangle
\otimes\left[  \left(  1-\frac{1}{2}\frac{\varepsilon^{2}}{\kappa^{2}}\right)
\left\vert 0\right\rangle _{c}-i\frac{\varepsilon}{\kappa}\left\vert
1\right\rangle _{c}\right]  . \label{rhossdark}
\end{equation}

On the other hand, if the system is in the $\left\vert G,0\right\rangle $
state, the weak probe field could \textit{a priori} induce two off-resonant
transitions: $\left\vert G,0\right\rangle \leftrightarrow\left\vert
-\right\rangle $ and $\left\vert G,0\right\rangle \leftrightarrow\left\vert
+\right\rangle $, with detuning between the frequencies of the probe and
atom-field system given by $\sqrt{2}g$. However, when $\omega_{p}=\omega$, the
probe field does not introduce any photon into the cavity in the stationary
regime, no matter the value of the atom field coupling $g$ (bellow we explain
this point in more detail). Then, when the system is in the $\left\vert
G,0\right\rangle $ state, the probe field is not able to introduce any
excitation in the system so that the steady state of the system is
\begin{equation}
\left\vert \psi\right\rangle _{ss}^{G}\approx\left\vert G,0\right\rangle .
\label{rhossground}
\end{equation}

Therefore, the normalized transmission of the cavity, $T=\left\langle a^{\dag
}a\right\rangle /\left(  \left\vert \varepsilon\right\vert /\kappa\right)
^{2}$, is useful to provide us information about the atomic steady state
(\ref{rhoss}), i.e., we have $T=1$ when the atoms are in the maximally
entangled state and $T=0$ when the atoms are in a separable state. So, in the
stationary regime, after applying a\ weak probe field on the system, the
transmission works out as a nondemolition measurement of the atomic state,
allowing us to know whether the system is in a maximally entangled state or
not. Besides, our system does not require a high efficiency photon detector
since a single click is enough to discriminate between the two atomic states
present in the steady state (\ref{rhoss}). If we are interested in preparing
an entangled state, we can simply monitor the transmission in the time
interval $\kappa/g^{2}\ll t<1/\gamma$: any click on the detector within this
time window projects the system in the maximally entangled state. If no click
is registered, then we must reset the system and start the experiment again.

\begin{figure}
\includegraphics[height=5.3444cm,width=6.4625cm]{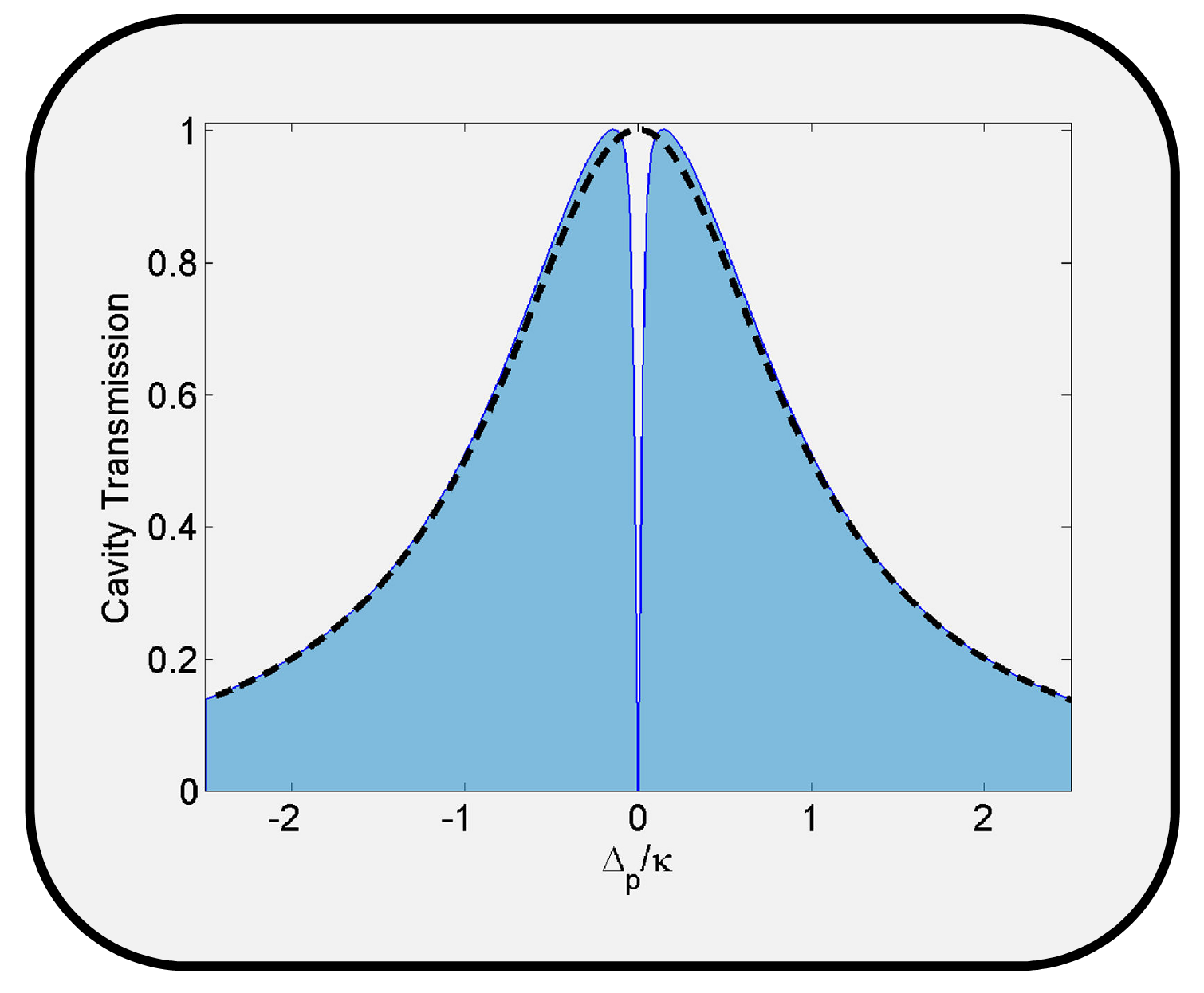}{}
\caption{(color online). Cavity transmission \textit{versus} detuning between
the probe and the cavity field considering $g=0.1\kappa$ and $\varepsilon
=0.1g$. We observe that, even for $\sqrt{2}g\lesssim\kappa/2$, the cavity
transmission is close to zero for $\Delta_{p}=0$ when $\rho_{ss}
\rightarrow\left\vert G,0\right\rangle $ (solid line). The dashed line
represents the cavity transmission when $\rho_{ss}\rightarrow\left\vert
D,0\right\rangle $ (empty cavity-like).}
\label{cavitytrans}
\end{figure}

The total transmission is expected to be maximum when the atoms are in the
dark state $\left\vert D\right\rangle $ because in this state the atomic
system is decoupled from the cavity mode so that the atom-field system behaves
as an empty cavity case ($g=0$). However, when the system is in the
$\left\vert G,0\right\rangle $ state, the transmission is expected to be zero
(or close to zero). The origin of this zero transmission could be in the
detuning between the weak probe field and the atom-field system: the two
transitions $\left\vert G,0\right\rangle \leftrightarrow\left\vert
-\right\rangle $ and $\left\vert G,0\right\rangle \leftrightarrow\left\vert
+\right\rangle $ are coupled by the probe field, but with detuning $-\sqrt
{2}g$ and $\sqrt{2}g$, respectively. As both states $\left\vert \pm
\right\rangle $ have decay rates $\Gamma_{\pm\rightarrow G,0}=\kappa/2$, one
can see that, for $\sqrt{2}g\gg$ $\kappa/2$, the probe field is very out of
resonance with the atom-field system and then it is expected an absorption
close to zero. If this is the case, one could argue that our system only works
in the strong coupling regime. However, our scheme is also valid for weak
atom-field coupling $g$, as the real reason why there is no transmission from
the cavity is that our system has two absorption channels $\left\vert
G,0\right\rangle \leftrightarrow\left\vert -\right\rangle $ and $\left\vert
G,0\right\rangle \leftrightarrow\left\vert +\right\rangle $ which
destructively interfere producing zero absorption in the resonant case
$\omega_{p}=\omega$, analogously to the phenomenon of electromagnetically
induced transparency \cite{EIT}. In our case the probe field is reflected by
the cavity mirror owing to this destructive interference \cite{EIT2}. The Fig.
\ref{cavitytrans} shows the cavity transmission as a function of the detuning
between the probe field and the cavity mode, $\Delta_{p}=\omega_{p}-\omega$,
considering the atom-field coupling $g=0.1\kappa$ and the strength of the
probe field $\varepsilon=0.1g$. Then, our scheme works out for any value of
$g$. However, the smallest the $g$ the longest the time to the system reach
the steady state since it is proportional to $\kappa/g^{2}$.%

\begin{figure}
\includegraphics[height=2.1741in,width=2.9897in]{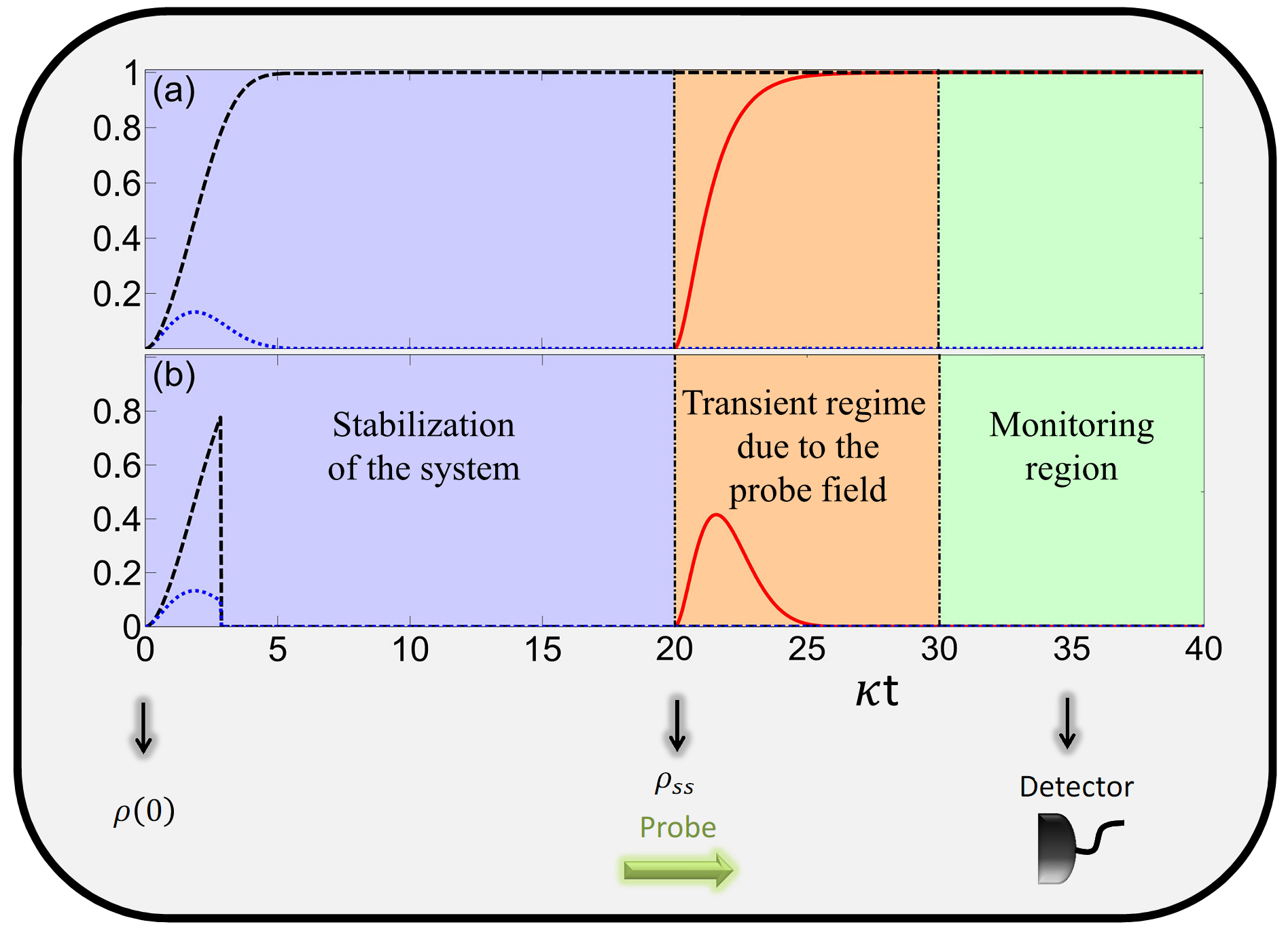}{}
\caption{(color online). Simulation of an experiment through the quantum jump
approach for $g=0.5\kappa$, $\varepsilon=0.05g$ and $\rho\left(  0\right)
=\left\vert \phi\left(  0\right)  \right\rangle \left\langle \phi\left(
0\right)  \right\vert $, with $\left\vert \phi\left(  0\right)  \right\rangle
=\left\vert g\right\rangle _{1}\otimes\left\vert e\right\rangle _{2}
\otimes\left\vert 0\right\rangle _{c}$. Time evolution of the mean number of
photons $\bar{n}=\left\langle a^{\dag}a\right\rangle $ (dotted line), the
concurrence $C$ (dashed line) and normalized transmission $T$ (solid line) for
a single trajectory when (a) $\rho_{ss}\rightarrow\left\vert D,0\right\rangle
$ and (b) $\rho_{ss}\rightarrow\left\vert G,0\right\rangle $.}
\label{quantumjump}
\end{figure}

To simulate an experiment, we employed numerical simulations using the quantum
jump approach (also called the quantum trajectories method) \cite{QJump}. It
is shown in Fig. \ref{quantumjump} a single trajectory simulating a single run
of an experiment for the case when $\rho_{ss}\rightarrow\left\vert
D,0\right\rangle $ [Fig. \ref{quantumjump}(a)] and when $\rho_{ss}
\rightarrow\left\vert G,0\right\rangle $ [Fig. \ref{quantumjump}(b)]. In these
simulations we used $g=0.5\kappa$, $\varepsilon=0.05g$ and $\rho\left(
0\right)  =\left\vert \phi\left(  0\right)  \right\rangle \left\langle
\phi\left(  0\right)  \right\vert $, with $\left\vert \phi\left(  0\right)
\right\rangle =\left\vert g\right\rangle _{1}\otimes\left\vert e\right\rangle
_{2}\otimes\left\vert 0\right\rangle _{c}$. Here, the quantification of the
degree of entanglement is done through the Wootters' concurrence $\left(
C\right)  $ \cite{Wooters,Conc}. As we can see in Fig. \ref{quantumjump}, when
the atoms are in the maximally entangled (separable) state, the transmission
of the probe field in the monitoring region is maximum (zero). This figure
also helps us to see the evolution of a single trajectory of the system: in
$t=0$ we have the preparation of the initial state $\rho\left(  0\right)  $,
followed by the stabilization of the system; then we switch on the probe
field, which requires a second stabilization time; finally we have the
monitoring region where the atomic state is nondestructively measured.

As we could see so far, to be able to generate the maximally entangled state
it is required that the initial atomic state $\rho\left(  0\right)  $ has a
nonzero projection on the dark state $\left\vert D\right\rangle $. This can be
done in different ways, for example: \textit{i)} if we are able to address the
atoms individually, then one can prepare the initial state\ $\left\vert
g\right\rangle _{1}\otimes\left\vert e\right\rangle _{2}\otimes\left\vert
0\right\rangle _{c}$; \textit{ii)} if we are not able to address the atoms
individually, then one can apply an incoherent field on both atoms
simultaneously so that one can prepare a completely mixed state $\rho\left(
0\right)  =\frac{\boldsymbol{1}_{a}}{4}\otimes\left\vert 0\right\rangle
_{c}\left\langle 0\right\vert $, with $\boldsymbol{1}_{a}$ being the identity
atomic matrix ($\boldsymbol{1}_{a}$ $=\left(  \left\vert g\right\rangle
\left\langle g\right\vert +\left\vert e\right\rangle \left\langle e\right\vert
\right)  _{1}\otimes\left(  \left\vert g\right\rangle \left\langle
g\right\vert +\left\vert e\right\rangle \left\langle e\right\vert \right)
_{2}$). In the first case, the projection on the dark state is $P=1/2$ while
in the second one $P=1/4$, which are the probabilities of preparing the atoms
in a maximally entangled state.

\textit{Direct measurement of the concurrence:} besides using our scheme as a
source to produce maximally entangled states, our scheme can also be used as a
direct method to measure the concurrence of the atoms. As we explained above,
for any initial state, the steady state of the atom-field system is given by
the Eq. (\ref{rhoss}), which is a mixture between a completely separable state
and a maximally entangled one. By applying a weak probe field, the steady
state turns out to be
\begin{equation}
\rho_{ss}\rightarrow\widetilde{\rho}_{ss}\approx\left(  1-P\right)  \left\vert
\psi\right\rangle _{ss}^{G}\left\langle \psi\right\vert +P\left\vert
\psi\right\rangle _{ss}^{D}\left\langle \psi\right\vert .
\end{equation}
with $\left\vert \psi\right\rangle _{ss}^{G}$ and $\left\vert \psi
\right\rangle _{ss}^{D}$ belonging to distinct subspaces. For this state, the
average transmission is $T\left(  \widetilde{\rho}_{ss}\right)  =P$. However,
the concurrence of the atomic state is also $C\left[  \text{Tr}_{c}\left(
\rho_{ss}\right)  \right]  =P$, where Tr$_{c}$ means the trace over the cavity
mode variables. Therefore, we see that the transmission of the atom-field
system $T\left(  \widetilde{\rho}_{ss}\right)  $ is exactly the degree of
entanglement (concurrence) between the two atoms. In this way, our scheme
works out as a direct method to measure of the concurrence of the atomic
steady state, without requiring any tomographic reconstruction of the atomic
density matrix.

In conclusion, we have shown a probabilistic scheme to prepare a maximally
entangled state between a pair of two-level atoms inside a leaking cavity, and
how to probe this atomic steady state, without perturbing it, via the cavity
transmission. From the point of view of a single run of the experiment, we
have seen that if the atomic system is in an entangled state, then there will
be a nonzero cavity transmission. On the other hand, if the system is in an
uncorrelated state, the cavity transmission goes to zero. In this way, a
single click in the detector is enough to ensure that the atoms are in an
maximally entangled state. We have also seen that our scheme works out as a
direct method to measure of the concurrence of the atomic steady state,
without requiring any tomographic reconstruction of the atomic density matrix.

\begin{acknowledgments}
The authors acknowledge the financial support from the Brazilian agencies
FAPESP, CNPq and Brazilian National Institute of Science and Technology for
Quantum Information (INCT-IQ).
\end{acknowledgments}

\end{document}